\documentclass[runningheads]{llncs}
\usepackage{amsmath}
\usepackage{float}
\usepackage[T1]{fontenc}
\usepackage{subcaption}

\usepackage{graphicx}
\begin{document}
\title{A Deep Generative Model of Neonatal Cortical Surface Development}
%
%
 \author{Abdulah Fawaz \inst{1} \and
 Logan Z. J. Williams \inst{1,2} \and
 A. David Edwards \inst{2,3,4} \and
 Emma Robinson \inst{1,2}  
 } 

 \authorrunning{A. Fawaz et al.}
 \institute{ Department of Biomedical Engineering, School of Biomedical Engineering and Imaging Sciences, King’s College London, London, United Kingdom 
 \and 
 Centre for the Developing Brain, School of Biomedical Engineering and Imaging Sciences,King's College London, London, United Kingdom
 \and
 Department for Forensic and Neurodevelopmental Sciences, Institute of Psychiatry, Psychology and Neuroscience, King’s College London, London, SE5 8AF, UK
 \and
 MRC Centre for Neurodevelopmental Disorders, King’s College London, London, SE1 1UL, UK
 }

\maketitle              
\begin{abstract}
The neonatal cortical surface is known to be affected by preterm birth, and the subsequent changes to cortical organisation have been associated with poorer neurodevelopmental outcomes. Deep Generative models have the potential to lead to clinically interpretable models of disease, but developing these on the cortical surface is challenging since established techniques for learning convolutional filters are inappropriate on non-flat topologies. To close this gap, we implement a surface-based CycleGAN using mixture model CNNs (MoNet) to translate sphericalised neonatal cortical surface features (curvature and T1w/T2w cortical myelin) between different stages of cortical maturity. Results show our method is able to reliably predict changes in individual patterns of cortical organisation at later stages of gestation, validated by comparison to longitudinal data; and translate appearance between preterm and term gestation ($>37$ weeks gestation), validated through comparison with a trained term/preterm classifier. Simulated differences in cortical maturation are consistent with observations in the literature.
\end{abstract}

\section{Introduction}

Deep Generative modelling presents enormous opportunities for medical imaging analysis: from image segmentation \cite{10.3389/fradi.2021.664444,costa2017end,yan2019domain,cirillo2020vox2vox}, registration \cite{yan2018adversarial}, motion correction and denoising \cite{ran2019denoising,yang2018low,latif2018automating,armanious2019retrospective}, to anomaly detection \cite{han2021madgan,zenati2018efficient,benson2020gan} and the development of clinically interpretable models of disease progression \cite{bass2020icam,bass2021icam,li2020dc}. 
Image-to-image translation is a type of generative modelling problem where images are transformed across domains, in a way that preserves their content. Common image-to-image translation tasks include transforming between imaging modalities \cite{hiasa2018cross,yang2018unpaired,welander2018generative} and to increase image resolution \cite{do2021mri,you2019ct,jiang2021fa}, though the methods used are general to any sets of imaging domains or classes, and can be applied to both 2D and 3D imaging data (see \cite{yi2019generative} for a full review of GANs in medical imaging). 

The application of deep learning on surfaces, including medical surfaces, has been hindered by the mathematical incompatibility of data structures with non-flat topologies to conventional approaches for convolutional filtering. In response, the field of geometric deep learning (gDL) has emerged to extend deep learning to domains such as graphs, surfaces and meshes \cite{Bronstein_review}. gDL is yet to produce a singular approach and numerous, often competing, tools and methods have been developed. 


In this paper, we aim to develop surface-to-surface translation models of the developing cortex sensitive to individual changes in cortical maturation. Such difference maps could act as imaging biomarkers of interest to clinicians, since the negative impact of preterm birth on neurodevelopmental outcome remains an area of active research \cite{morel2021automated,boardman2020invited}. This task is too challenging for traditional imaging analysis methods due to the significant variation in functional and structural organisation of human cortices, even amongst healthy populations. The heterogeneity is present in neonatal cortical surface patterns from an early stage, which then undergo rapid and complex cortical maturation during development, which are further obfuscated by the impact of preterm birth. These factors make this task a fitting domain for the use of gDL, and gDL models have already been shown to achieve state of the art performance for neurodevelopmental phenotype regression and on cortical segmentation, when benchmarked against classical methods \cite{sphericalunet,Fawaz2021.12.01.470730,vosylius2020geometric,dahan2021improving,williams2021geometric}. To achieve this, we adapt the CycleGAN framework \cite{cyclegan-original} with methods from gDL for use on surfaces. 



\section{Methods}

The CycleGAN framework \cite{cyclegan-original} is a well-established method for unsupervised bidirectional image-to-image translation across two domains. This is achieved with a pair of generators that learn mappings from a source domain to a target domain, trained adversarially to a corresponding pair of discriminators that differentiate between real and synthetic images. Traditionally, these generators and discriminators are Euclidean CNNs but in this work, we retain the architecture but replace the convolutions with a surface-compatible convolution operation from the MoNet model \cite{monet}. CycleGANs produce realistic images and do not require paired examples, which can be practically difficult to acquire in medical imaging. These fit in with our overall aim of generating realistic, subject-specific cross-domain difference maps that can identify individual deviations from normal morphological development. 

\subsection{Model Architecture}
The following adaptions were made to the original CycleGAN \cite{cyclegan-original} to facilitate application on surfaces:

\subsubsection{Geometric Convolutions:} All surface convolutions are implemented using MoNet Gaussian Mixture Model convolutions (GMMConv)  \cite{monet}:
\begin{equation}
\begin{split}
    &(f\star g)(x) = \sum_{j=1}^J g_j D_j(x)f \\
   & D_j(x)f = \sum_{y\in \mathcal{N}(x)} w_j(\mathbf{u}(x,y))f(y), \quad j=1,\ldots, J \label{eq:patch}
\end{split}
\end{equation}
where $f,g$ are the input and filter respectively, $D(x)f$ is a parametrisable patch operator that determines how the data is extracted from the surface before being weighted by the filter. Its general form is given in equation \eqref{eq:patch}, with $x$ a point on the surface, $y$ a point in the defined neighbourhood of $x$, and $w$ a kernel applied to the pseudo-coordinates $u(x,y)$ that define how the neighbourhood of an individual point is weighted relative to its neighbours on the surface. In MoNet, the weighting function $w$ takes a Gaussian form:
\begin{equation}
    w_j(\textbf{u}) = \exp(-\frac{1}{2}(\mathbf{u}-\mathbf{\mu}_j)^T \Sigma_{j}^{-1} (\mathbf{u}-\mathbf{\mu}_j))
\end{equation}
where $\Sigma_j$ and $\mu_j$ are learnable $d\times d$ and $d\times 1$ covariance matrix and mean vector of a Gaussian kernel, respectively. The covariances are further restricted to diagonal form.  In this paper, Gaussian kernels are defined using a two dimensional (J=2) local pseudo-coordinate system defined as the vector difference between neighbouring points on the surface in polar coordinates. This is pre-computed.


\subsubsection{Icospheric Pooling:} 
\begin{figure}[t]
\centering
\includegraphics[width=\textwidth]{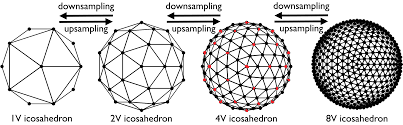}
\caption{Icosahedrons can be efficiently up and downsampled to different resolutions. New points are generated by barycentric interpolation of existing points. Our downsampling process simply keeps the points of the lower resolution icosphere while discarding the rest (e.g to downsample from 4V to 2V, keep the points shown in red). Our upsampling procedure mimics the upsampling of the icospheres but with the new points added as an average of the existing points within its neighbourhood.}
\label{fig:downsampling}
\end{figure}

Since GMMConv has no transposed convolution, and does not reduce the dimensionality of the data, we require pooling and unpooling layers adapted for the surface. This can be challenging for general surfaces, but icospheres of different resolutions may be generated from each other by iterative barycentric interpolation (Fig \ref{fig:downsampling}). This allows us to define a native down/upsampling method on the icosphere; the former as a direct downsample of the data by including only points of the lower icosphere (e.g. the red points in figure \ref{fig:downsampling}), and the latter as a mean-unpool operation where the new points are the average of their direct neighbours during the upsampling process. Skip connections are also added to make the generator a U-net.

\subsubsection{Training} The network is optimised following the protocol of a standard CycleGAN \cite{cyclegan-original}, which learns mappings $G_X: X\rightarrow Y$ and $G_Y: Y\rightarrow X$, through adversarial training. Unpaired images $\{x,y\}$ are input through each of the Generators, G, to produce synthetic images $\{\hat{x}_X, \hat{y}_X\}$ and $\{\hat{x}_Y, \hat{y}_Y\}$ from generators $G_X$ and $G_Y$ respectively. The identity loss penalises each generator for changing an image already in its domain i.e it forces $\hat{x}_X \approx x$, and $\hat{y}_Y \approx y$. The discriminators $D_x$ and $D_Y$ attempt to classify the true from synthetic images of their respective classes. Finally, the synthetic output of one generator is fed into the other $X\rightarrow Y\rightarrow X$ in order to generate a recovered image. The cycle loss penalises each generator for failing to reproduce the original image during this cycle.

We use an L1 loss for the cycle and identity losses, and an MSE loss for the adversarial loss. All networks are trained with an Adam optimizer with learning rate$=0.002$, betas$=[0.9, 0.99]$. We set the real and false labels to 0.9 and 0.1 to help avoid discriminator overfitting. Networks were trained for 250 epochs.

The data is augmented with non-linear warps to represent realistic variations. These were generated offline by creating surface meshes warped by very small random displacements in the vertices, and resampling the original data onto these warped surface meshes by barycentric interpolation \cite{Fawaz2021.12.01.470730}. 100 warped counterparts are created for each image in our dataset. Each imaging modality was normalised to between 0 and 1 across the dataset.

\section{Experimental Methods and Results}

\subsection{Data}
All of the data used are part of the third release of the publically available Developing Human Connection Project (dHCP) dataset, consisting of cortical surface meshes and metrics, derived from T1-weighted and T2-weighted MRI images, processed and registered as described in \cite{makropoulos2018developing} and references therein \cite{schuh2017deformable,hughes2017dedicated,kuklisova2011dynamic,kuklisova2012reconstruction}.
We utilise left cortical hemispheres scanned between 25 and 45 weeks post-menstrual age (PMA) with myelination and cortical curvature metrics initially registered to a 32k-vertex sphere using Multimodal Surface Matching (MSM) \cite{robinson2014msm,robinson2018multimodal} and upsampled to a regular icosphere. The data set contains 419 neonates born at term ($>=$37 weeks gestational age (GA) ) and 161 born preterm ($<$37 weeks gestational age), where 45 preterm subjects were scanned twice, once at birth and once at term-equivalent age (TEA), giving a total of 625 scans. 

\begin{figure}
\centering
\begin{subfigure}{0.5\textwidth}
   \includegraphics[width=1\linewidth]{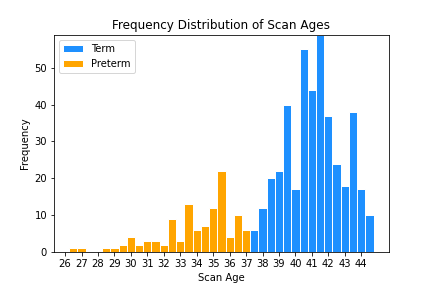}
   \caption{}
   \label{fig:dataset-sa} 
\end{subfigure}%
\begin{subfigure}{0.5\textwidth}
   \includegraphics[width=1\linewidth]{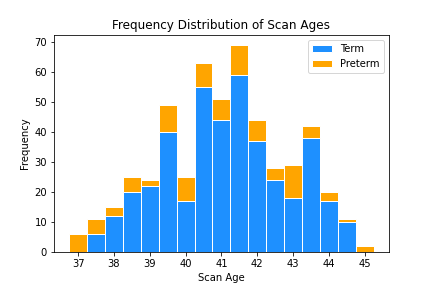}
   \caption{}
   \label{fig:dataset-ba}
\end{subfigure}
\caption{A bar chart showing the distribution of subjects' PMA at Scan in weeks for (a) Experiment 1 and (b) Experiment 2.}
\end{figure}

\subsection{Experiment 1: Cortical Maturity}
In this experiment, the objective was to simulate healthy cortical maturation; thus the two domains to be translated between were: A) preterm infants' first scans (PMA $<$37 weeks); and B) healthy term controls. We exclude the later scans of preterm subjects to avoid modelling atypical cortical maturation, giving a total of 530 subjects: 419 term/111 preterm (as shown in figure \ref{fig:dataset-sa}). 

A spatio-temporal atlas of cortical maturation maps is available for the dHCP data  \cite{bozek2018construction}, and shows a general increase in myelination around the brain, but especially strongly along the central sulcus (somatosensory cortex), the posterior portion of the lateral fissure (auditory cortex), and at higher scan ages, around the calcarine sulcus (visual cortex). By 28 weeks, all the primary sulci are formed and there are already heterogeneities in cortical structure. These sulci are initially smooth but become significantly more folded during development, and the cortical surface further increases in complexity by the emergence of tertiary sulci. Our model must be able to preserve and extend the existing folding patterns of an immature brain in its predictions.

The presence of longitudinal data in the form of preterm subjects' second scans allows us to quantitatively evaluate our model's predictions by comparing the actual appearance of a preterm subject's later scan to the model's matured prediction based on the same subjects first scan. 
We analyse these both quantitatively, via the image similarity metric Peak Signal to Noise Ratio (PSNR), and qualitatively by visual comparison. We compare the images as a whole and by individual modality (myelination/curvature). If the model represents cortical maturation, the model's synthetic predictions would be expected to be more similar to the later second scans than the original first scans. However, we would expect some differences due to variation around the age at scan and the impact of preterm birth on cortical development.
\begin{figure}
\centering
\begin{subfigure}{\textwidth}
   \includegraphics[width=1\linewidth]{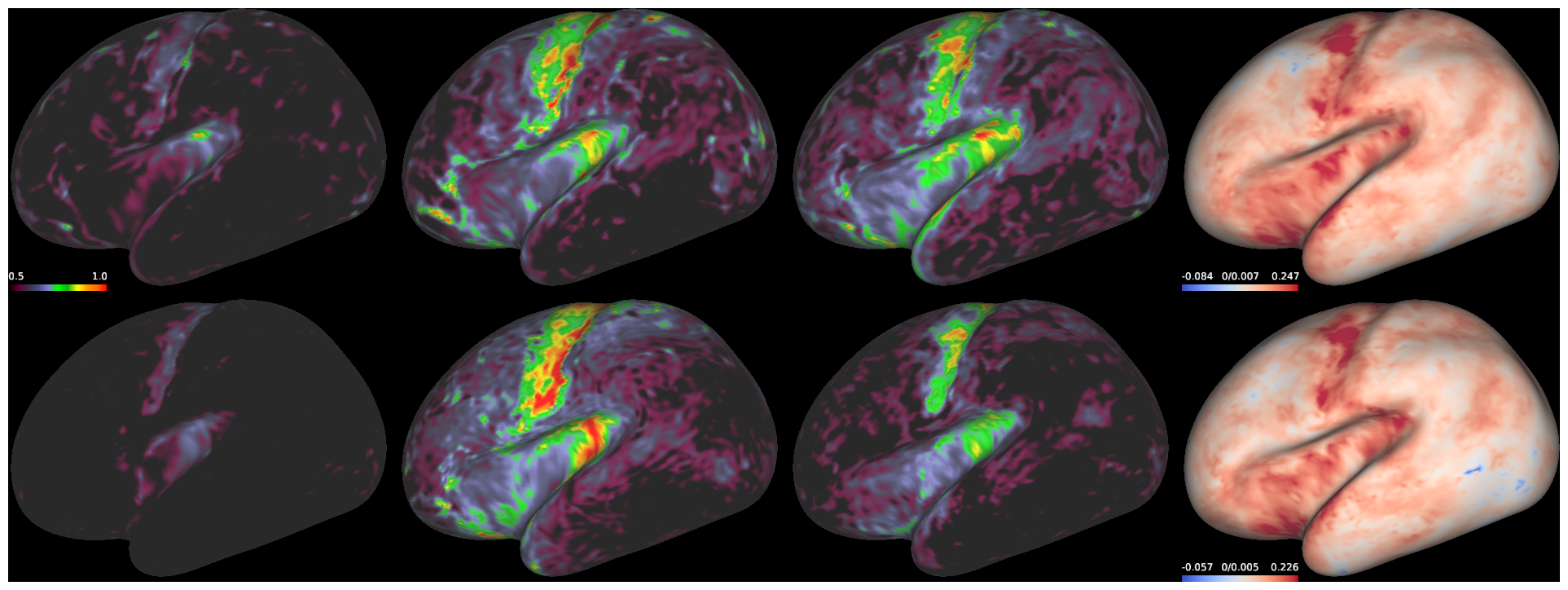}
   \caption{}
   \label{sa_myelination_low-high} 
\end{subfigure}

\begin{subfigure}{\textwidth}
   \includegraphics[width=1\linewidth]{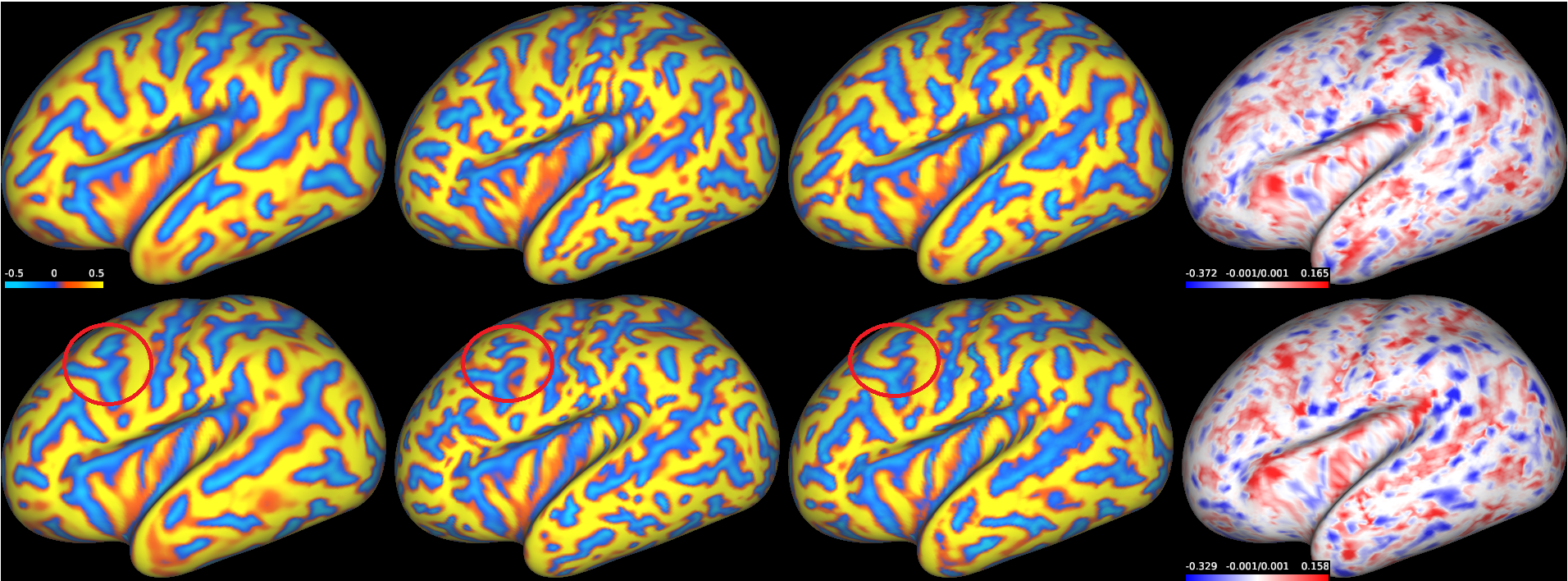}
   \caption{}
   \label{sa_curvature_low-high}
\end{subfigure}

\begin{subfigure}{\textwidth}
   \includegraphics[width=1\linewidth]{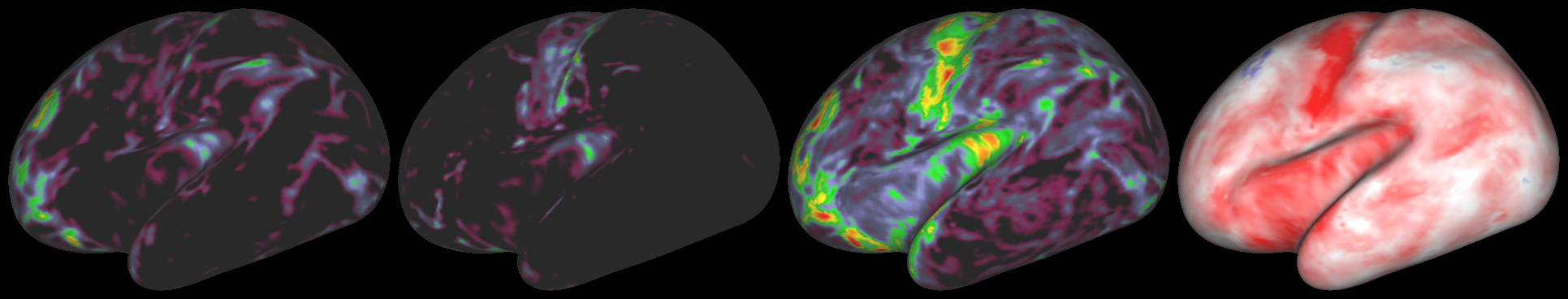}
   \caption{}
   \label{fig:too-myelin}
\end{subfigure}
\caption{Low to high PMA translation for the (a) myelination and (b) curvature modalities of two different subjects. The synthetic aged model predictions (column 3) are closer to the later second scans (column 2) than the original first scans (column 1). The fourth column gives the difference map between model input and model output. The red circles indicate a region of folding where the model correctly predicted a change in topology as the sulcus was split into two during development. (c) Myelination prediction for a subject with low PMA (39 weeks) at second scan showing the relative immaturity of the second scan myelination compared to the model prediction, resulting in a higher image similarity to the original first scan than to the model prediction.}
\end{figure}

Figures \ref{sa_myelination_low-high} and \ref{sa_curvature_low-high} show the predicted surface to surface translation for low-to-high PMA of two different subjects' myelination and curvature respectively, displaying only the lateral view for clarity. The first, second and third columns show, in order, the original first scans, the true second scans and the synthetic scans produced by the model. The final column is a difference map between the model input and the model output. Each row shows a different subject and it can be seen that the model correctly predicts the broad structural changes expected in both myelination and curvature for both subjects. The relative homogeneity of myelination makes it easy to see that the model has correctly learned to simulate subject development. 

Changes in curvature are more subtle and display greater heterogeneity across subjects and so are the better benchmark of performance. The preservation of cortical folding patterns is a key factor required for proper modelling of curvature and we see that our model does this well for both subjects despite significant topological differences between the two. Further, we observe that the model has added significant branching to most sulci, particularly the superior temporal sulcus which increases greatly in folding. The model has created new smaller folds in the temporal lobe, the parietal lobe and the frontal lobe and, extended existing sulci throughout. In the second row, the model even correctly predicts a change in the topology of a sulcus in the frontal lobe, which is split by a gyrus as a it develops (shown by the red circle). 

\begin{figure}[h]
\centering
\begin{subfigure}{0.6\textwidth}
   \includegraphics[width=1\linewidth]{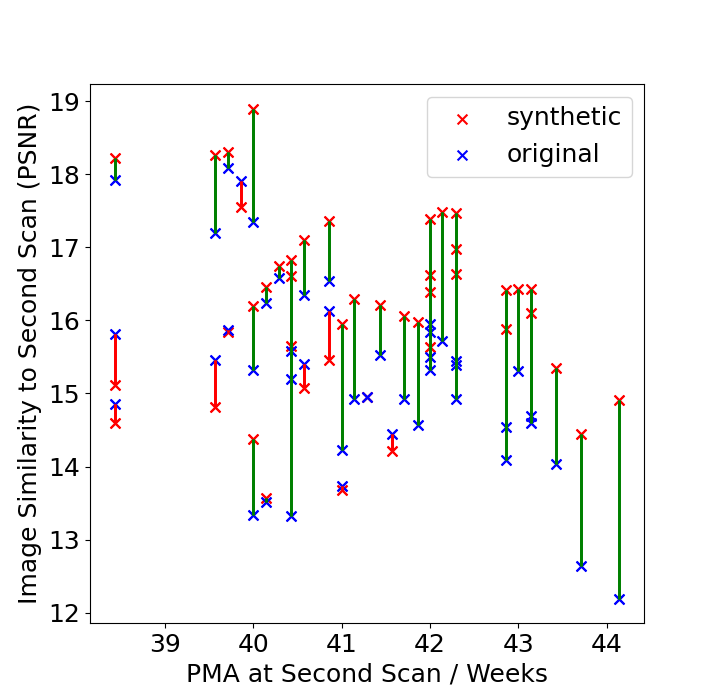}
   \caption{}
   \label{fig:scatterplot} 
\end{subfigure}
\begin{subfigure}{0.43\textwidth}
   \includegraphics[width=1\linewidth]{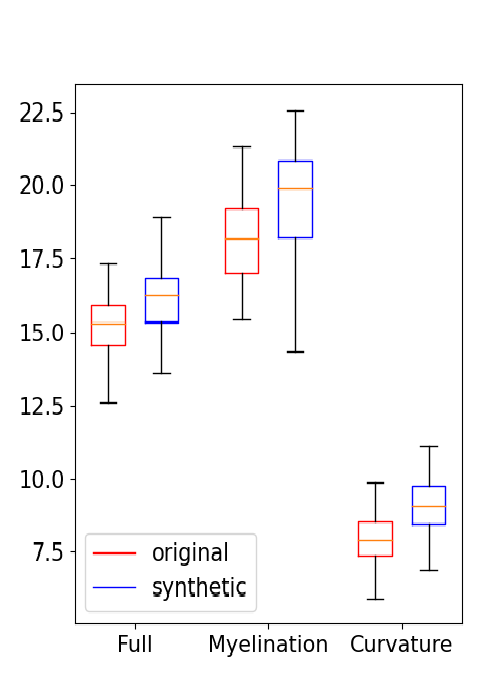}
   \caption{}
   \label{fig:boxplots}
\end{subfigure}

\caption{(a) Image similarity of original first scans vs synthetic aged images compared to the same subject's second scans, given by Peak Signal-to-Noise Ratio (PSNR). PSNR is plotted against PMA at second scan in weeks. Green lines indicate that the synthetic images are closer to the second scan than original first scans, and red lines indicate a reduction in image similarity. (b) Box plots comparing the range of image similarities as measured by PSNR of original first scans and synthetic aged scans to second scans, split by the full scan (both modalities), then by each modality individually.}
\end{figure}

Figure \ref{fig:scatterplot} is a scatterplot showing the quantified similarity between the second scans, which acts as a fuzzy 'ground truth', and the model's synthetic aged scans as compared to the first original scan, plotted against PMA at second scan. The results show that, in almost every case, the model predictions demonstrate a significant increase in similarity to the second scan compared to the original. The exceptions are mostly confined to subjects with second scans at low PMA. The results are summarised in the boxplots in figure \ref{fig:boxplots}, where the synthetic images are of greater similarity than the originals over both modalities individually and combined. We observe that, whilst synthetically generated curvature is consistently improved, there is a greater variation in myelination with a tail of synthetic images with low image similarity to the second scans. Visual inspection reveals that this discrepancy is responsible for the reduction in image similarity of synthetic images for low PMA second scans due to the model's predictions being significantly more mature than both the first and second scans in these cases, as seen in figure \ref{fig:too-myelin} where the predicted myelination is significantly greater than found in the second scan, but still consistent with expected development. This discrepancy is a consequence of the discontinuous nature of cycleGAN translations.

\subsection{Experiment 2: Prematurity}
In this experiment, the aim is to translate between preterms' second scan and healthy term controls, in order to predict the impact of preterm birth on each individual's cortical maturation. We exclude preterm first scans (PMA at scan $<37$ weeks) to only model the effect of preterm birth at term equivalent ages (TEA), leaving a total of 514 neonatal subjects (419 term/95 preterm). A plot of the cohort used is shown in figure \ref{fig:dataset-ba}. 

In this case, there can be no ground truth examples scanned at multiple birth ages, so to validate we train a classifier with same architecture as the discriminator to predict prematurity and apply it to our model's predictions. Our classifier obtained an accuracy of 94\% on the raw dataset, but our cycleGAN model's synthetic images had an overall 78\% success rate in fooling the classifier, and in the other 22\% of cases reduced the classifiers confidence by a significant margin.

\begin{figure}[h]
\includegraphics[width=\textwidth]{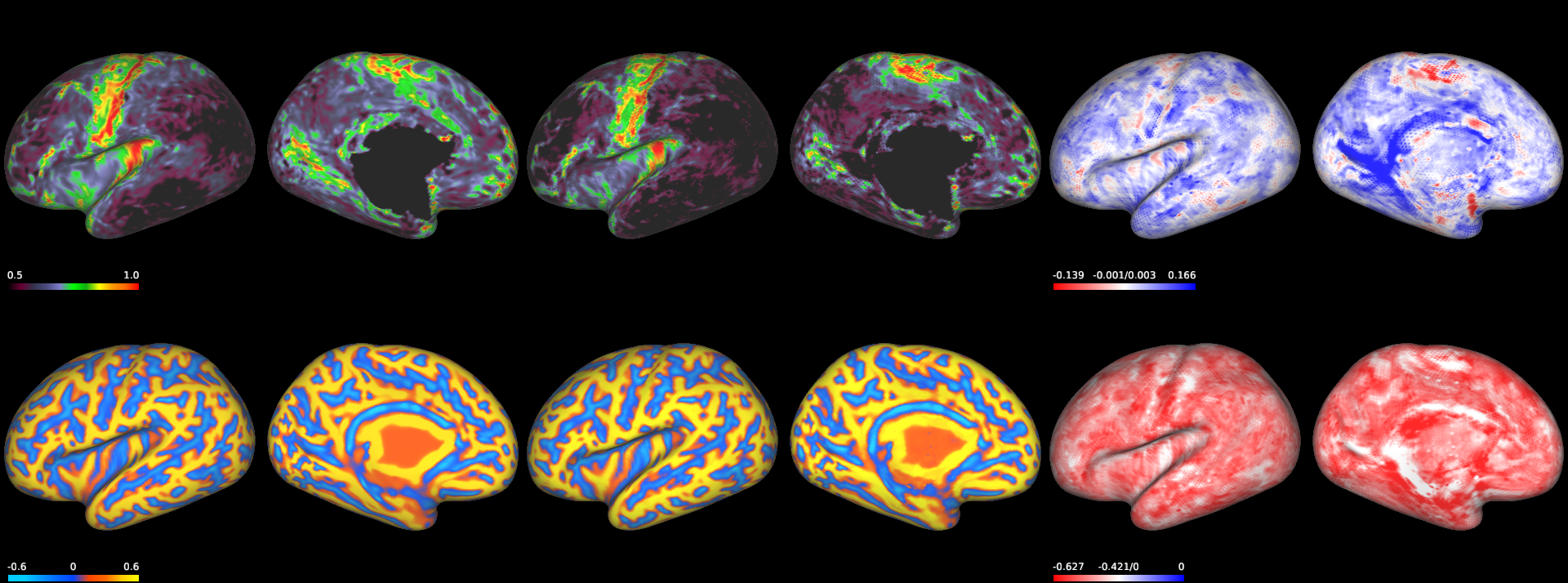}
\caption{Original Term (columns 1-2) to Synthetic Premature (columns 3-4) image translation with difference map (columns 5-6). Subject's true GA at birth is 40 weeks and was scanned at 44 weeks PMA. Classifier predicts $P_{\textrm{prem}}=0.22$ for the original term image and $P_{\textrm{prem}}=0.96$ for the synthetic premature image. } \label{birth_age_25}
\end{figure}

Figure \ref{birth_age_25} shows a subject (true GA at birth 40 weeks, PMA at scan 44 weeks) (column 1) that has been translated from term age to a synthetic preterm age (column 2). The image fools our classifier which predicts prematurity with $p=0.96$ for the synthetic preterm and $p=0.22$ for the original term subject. The difference map (column 3) shows that myelination is decreased overall but remains broadly unchanged in structure. The folding patterns are completely unchanged but there is a noticeable increase in overall curvature, and it can be seen from the difference map that increases in curvature are more pronounced along the gyri (folds), not the sulci. Comparing these observations to existing work, we note that increased overall cortical curvature has previously been associated with preterm birth when compared to age-equivalent fetuses \cite{lefevre2015developmental}, and when compared with normal term infants \cite{makropoulos2016regional}, with curvature especially high across a number of gyri. Increased curvature has also been implicated as a prognostic biomarker of adverse neurodevelopment \cite{kline2020early}. Shimony et al \cite{shimony2016comparison} also found increased overall curvature in preterm neonates compared to term neonates, with increased localised curvature around the gyri, but could not positively determine that the metric was predictive of preterm birth due to the differences in folding patterns between subjects obfuscating the measure - an issue circumvented here as we compare like-for-like subjects through our cycleGAN.

\section{Conclusion}
By integrating MoNet into a CycleGAN model, we developed a generative surface-to-surface translation model of cortical maturation that shows accurate, realistic, subject-specific predictions of future myelination and curvature, validated on longitudinal ground truth data. We also developed a model that transformed surfaces between term/preterm classes, producing outputs that fooled a pretrained classifier and showed structural differences in line with what has been observed in the literature.
The primary issue with our models was that the CycleGAN architecture is limited to transformation between discrete classes when this is a fundamentally a problem of continuous interpolation. There are a number of different generative models that do allow continuous representation that would be able to interpolate smoothly between scan ages, with which gDL could be adapted. One class of methods utilise variational encodings such as variational autoencoders (VAEs) \cite{kingma2013auto}, and VAE-GANs \cite{larsen2016autoencoding}, although the authors found that attempts to adapt these methods to surface domains were unsuccessful, with models unable to capture individual cortical folding variation and collapsing to group averages. A more powerful example of this applied to volumetric cortical data is the iCAM architecture \cite{bass2021icam}, which encodes separate variational disentangled spaces for content and age, which may be more amenable to adaptation with gDL. 
There are further alternatives that utilise direct conditioning on latent variables such as conditional VAEs and conditional GANs, but these too are more commonly associated with conditioning on classes, not continuous variables. Again more complex variations of these models have been successfully applied to the volumetric brain \cite{xia2021learning} that may be successfully adapted to the surface.

%
%
%
\bibliographystyle{splncs04}
\bibliography{bibliography}

\end{document}